# Neocortical Dynamics at Multiple Scales:
## EEG Standing Waves, Statistical Mechanics, and Physical Analogs

Lester Ingber[1] and Paul L. Nunez[2]

ABSTRACT: The dynamic behavior of scalp potentials (EEG) is apparently due to some combination of global and local processes with important top-down and bottom-up interactions across spatial scales. In treating global mechanisms, we stress the importance of myelinated axon propagation delays and periodic boundary conditions in the cortical-white matter system, which is topologically close to a spherical shell. By contrast, the proposed local mechanisms are multiscale interactions between cortical columns via short-ranged non-myelinated fibers. A mechanical model consisting of a stretched string with attached nonlinear springs demonstrates the general idea. The string produces standing waves analogous to large-scale coherence EEG observed in some brain states. The attached springs are analogous to the smaller (mesoscopic) scale columnar dynamics. Generally, we expect string displacement and EEG at all scales to result from both global and local phenomena. A statistical mechanics of neocortical interactions (SMNI) calculates oscillatory behavior consistent with typical EEG, within columns, between neighboring columns via short-ranged non-myelinated fibers, across cortical regions via myelinated fibers, and also derive a string equation consistent with the global EEG model.

## 1. Introduction

What makes human brains so special? How do they differ from hearts, livers, and other organs? All organ systems are enormously complicated structures, able to repair themselves and make detailed responses to external control by chemical or electrical input. Yet, only brains yield the amazing phenomenon of consciousness (Nunez, 2010). Complex adaptive systems, for which human brains provide the most prominent examples, are composed of smaller parts interacting both within and across spatial scales. They typically exhibit emergent behavior not obviously predictable from knowledge of the individual parts and have the added capacity to learn from experience and change their global behaviors by means of feedback processes. Other examples include stock markets, ecosystems, and all living systems.

Several general features distinguish human brains from other organs, including the hallmark of richer hierarchical (or multi-scale) interactions. In contrast to simple cognitive "theories," this paper explicitly acknowledges brains as highly complex adaptive systems, emphasizing the critical contribution of cross scale interactions to their dynamic behaviors. In order to minimize communication barriers due to the complicated mathematics, several analog systems from disparate fields are employed. Neuroscientists are typically skeptical of brain analogs, typically for good reason; however, we are *not* claiming that brains are actually just like stretched strings, social systems, quantum structures, resonant cavities, hot plasmas, disordered solids, chaotic fluids, or any other non-neural system. Rather, we suggest that each of these systems may exhibit behavior similar to brain dynamics observed under restricted experimental conditions, including the spatial scale of observation. The multiple analogs then facilitate development of complementary models of brain reality.

In many complex systems, as spatial-temporal scales of observation are increased, new phenomena become evident by virtue of synergistic interactions among smaller-scale entities, which serve to explain data, typically in a mathematically aesthetic fashion (Haken, 1983; Nicolis & Prigogine, 1973). For example, in the classical thermodynamics of equilibrium systems, it is possible to transition from microscopic molecular scales to macroscopic scales and employ the macroscopic variable temperature to describe the average kinetic energy of microscopic molecular activity. Many complex systems, however, operate in non-equilibrium states, being driven by nonlinear and stochastic interactions. For such systems, classical thermodynamics typically does not apply (Ma, 1985). For example, the description of weather and ocean patterns, which includes important features such as turbulence, rely on semi-phenomenological mesoscopic models, in agreement with molecular theories but not capable of being rigorously derived from them. Phase transitions in magnetic systems and many systems similarly modeled (Ma, 1976; Wilson, 1979; Wilson & Kogurt, 1974) require careful treatment of a continuum of scales near critical points. In general, rather than having a general theory of non-equilibrium nonlinear process, several overlapping approaches are employed, typically geared to classes of systems and often expanding on nonlinear treatments of stochastic systems (Gardiner, 1983; Haken, 1983; Kubo, Matsuo & Kitahara, 1973; Nicolis & Prigogine, 1973; van Kampen, 1981).

[1]Lester Ingber
Lester Ingber Research, Ashland, OR
ingber@alumni.caltech.edu
http://www.ingber.com

[2]Paul L. Nunez
Tulane University and Cognitive Dissonance, LLC, New Orleans, LA
pnunez@tulane.edu
http://www.linkedin.com/in/paulnunez

Given this general outline of complex systems, it should not be surprising that human brains support many phenomena arising at different spatial-temporal scales. We can then study macroscopic neocortical phenomena such as electroencephalography (EEG) by appealing to a chain of arguments dealing with overlapping microscopic and mesoscopic scales. Such work is detailed in a series of papers presenting a theory of statistical mechanics of neocortical interactions (Ingber, 1981; Ingber, 1982; Ingber, 1995a; Ingber & Nunez, 1990). This approach permits us to develop EEG and other models of dynamic processes whose variables and parameters are closely identified with ensembles of synaptic and neuronal interactions. The mathematical formalism supporting this approach has only recently been made possible by developments in mathematical physics since the late 1970s, in the field of nonlinear non-equilibrium statistical mechanics. The origins of this theory are in quantum and gravitational field theory.

## 2. EEG and other experimental data

The ultimate test of any brain model is experiment, and different kinds of brain data are available at different spatial and temporal scales. Structural or static imaging is accomplished with computed tomography (CT) or magnetic resonance imaging (MRI). The label "static imaging" indicates changes on yearly time scales in healthy brains or perhaps weeks or months in the case of growing tumors. By contrast, intermediate time-scale methods like functional magnetic resonance imaging (fMRI) and positron emission tomography (PET) track brain changes over seconds or minutes. Still more rapid dynamic measures are electroencephalography (EEG) and magnetoencephalography (MEG), which operate on millisecond time scales, providing dynamic images faster than the speed of thought. The "bad" news is that EEG spatial resolution is quite coarse; nevertheless, EEG provides most of the existing data on neocortical dynamic behavior and its relation to cognitive events in humans. Thus, we focus on EEG a record of the oscillations of brain electric potential recorded from electrodes on the human scalp.

EEG allows for accurate identification of distinct sleep stages, depth of anesthesia, seizures and other neurological disorders. It also reveals robust correlations with cognitive processes occurring during mental calculations, working memory and selective attention. Scientists are now so accustomed to these EEG correlations with brain state that they may forget just how remarkable they are. The scalp EEG provides very large-scale and robust measures of neocortical dynamic function. A single electrode yields estimates of synaptic action averaged over tissue masses containing between roughly 100 million and 1 billion neurons. The space averaging of brain potentials resulting from extra-cranial recording is a fortuitous data reduction process forced by current spreading in the head volume conductor. Much more detailed local information may be obtained from intracranial recordings in animals and epileptic patients. However, intracranial electrodes implanted in living brains provide only very sparse spatial coverage, thereby failing to record the "big picture" of brain function. Furthermore, the dynamic behavior of intracranial recordings depends fundamentally on measurement scale, determined mostly by electrode size. Different electrode sizes and locations can result in substantial differences in recorded dynamic behavior, including frequency content and coherence. Thus, in practice, intracranial data provide different information, not more information, than is obtained from the scalp (Nunez & Srinivasan, 2006a).

We expect brain electrical dynamics to vary substantially across spatial scales. Although cognitive scientists and clinicians have reason to be partly satisfied with the very low spatial resolution obtained from scalp EEG data, explorations of new EEG methods to provide somewhat higher spatial resolution continue. A reasonable goal is to record averages over "only" 10 million neurons at the one-cm scale in order to extract more details of the spatial patterns correlated with cognition and behavior. This resolution is close to the theoretical limit of spatial resolution caused by the physical separation of sensor and brain current sources. MEG spatial resolution is also quite limited because its sensor coils are at least three times further from dominant brain sources than EEG electrodes (Nunez & Srinivasan, 2006a). Scalp data are largely independent of electrode size because scalp potentials are severely space-averaged by volume conduction between brain and scalp. Intracranial recordings provide much smaller scale measures of neocortical dynamics, with scale depending on the electrode size, which may vary over four or five orders of magnitude in various practices of electrophysiology. A mixture of coherent and incoherent sources generates the small and intermediate scale intracranial data. Generally, the smaller the scale of intracranial potentials, the lower the expected contribution from coherent sources and the larger the expected differences from scalp EEG. That is, scalp data are due mostly to sources coherent at the scale of at least several centimeters with special geometries that encourage the superposition of potentials generated by many local sources.

In practice, intracranial EEG may be uncorrelated or only weakly correlated with cognition and behavior. The information content in such recordings is limited by sparse spatial sampling and scale-dependent dynamics. Furthermore, most intracranial EEG data are recorded in lower mammals; extrapolation to humans involves additional issues. Thus, higher brain function in humans is more easily observed at large scales. Scientists interested in higher brain function are fortunate in this respect. The technical and ethical limitations of human intracranial recording force us to emphasize scalp recordings. These extra-cranial recordings provide estimates of synaptic action at the large scales closely related to cognition and behavior. Thus, EEG provides a window on the mind, albeit one that is often clouded by technical and other limitations.

## 3.  Possible physiological bases for EEG

Since the first human recording in the early 1920s the physiological bases for the wide variety of rhythmic EEG activity, a proverbial "spectral zoo," has been somewhat of a mystery. In particular, human alpha rhythms, which are quite robust in wide awake (but relaxed) subjects with closed eyes, may be recorded over nearly all of the upper scalp or cortex and have preferred frequencies near 10 Hz. Given any unknown physical or biological system that produces oscillations at some preferred (or resonant) frequency $f = \omega/2\pi$, one of the first questions a scientist might ask concerns the origin of the implied underlying time delay $\tau$ roughly estimated as

$$\tau \sim \omega^{-1} \tag{1}$$

The implied physiological time scales for the most robust human EEG rhythms (1 to 15 Hz) are $\tau = 160$ -160 ms. How does this delay range compare with mammalian physiology? Whereas early studies of membrane time constants in mammalian cortex were very short, typically less than 10 ms, more modern studies with improved recording methods report the wide range 20 -100 ms (Koch, Rapp & Segev, 1996). But apparently in voltage-gated channels, the effective time constant becomes a "dynamical parameter" that depends on both membrane voltage and on time, thus genuine time constants are not really "constant." Koch et al. argue that the voltage response to very brief synaptic inputs is essentially independent of the classically defined time constant, which typically provides *overestimates* of the response time of neurons. In summary, these studies suggest that while synaptic delays (PSP rise and decay times) lie in a general range (within a factor of perhaps five or ten) that might account for dominant EEG frequencies, claims of close agreement between the *details* of observed EEG spectra and dynamic theories based on membrane time constants are not credible. Model parameters can be chosen to "match" favored EEG data sets, which, in any case, can vary widely between individuals and brain states.

By contrast to these "local" delays at the single neuron level, axonal ("global" delays along the longest white matter (corticocortical) fibers between anterior and posterior regions are estimated to be roughly in the 30 ms range in humans (Nunez, 1995). Such global delays depend on axon length distribution and axon propagation speed; thus they are expected to be much shorter in smaller mammalian brains if axon diameters (or propagation speed) are unchanged across species. To complicate matters, creation of serial connections between cell assemblies can apparently modify both local (PSP) and global (axon) characteristic delay times. While both local and global delays appear to be in a general range favorable for EEG production, this semi quantitative observation tells us little about the physiological mechanisms responsible for "special frequencies" like the narrow band human alpha rhythms or gamma oscillations (~40 Hz), the latter recorded mostly from inside the craniums of humans and lower mammals. Neither local theories (based on PSP rise and decay times) nor global theories (based on axon delays) can honestly claim close agreement with EEG data based *only* on predicted EEG spectral properties; the underlying physiological parameters (e.g., time constants and axonal delays) are not known with sufficient accuracy to make such claims credible. While PN has suggested that the parameters of the global standing wave theory appear to be known more accurately than local parameters, others may disagree. Nevertheless, we *can* agree to search for qualitative and semi quantitative connections between theory and EEG experiments that do not require precise physiological parameter knowledge.

The general idea of standing EEG waves that PN first proposed (Nunez, 1972; Nunez, 1974) was based on a very simple idea. Any kind of weakly damped, non-dispersive wave phenomenon propagating in a medium with characteristic speed $v$ can be expected to form standing waves due to wave interference that depends on the system's size and shape (the boundary conditions). Such phenomena occur, for example, in violin and piano strings and many other vibrating systems. Whereas waves in strings and flutes are reflected from boundaries, waves in closed systems like spherical shells or tori interfere because of periodic boundary conditions causing waves traveling in opposing directions to meet and combine. As a result of this interference, preferred (resonant) frequencies persist in such systems. Examples of standing waves in spherical geometry include the quantum wavefunction of the hydrogen atom (both radial and tangential waves) and the Schumann resonances of electromagnetic waves in the spherical shell formed by the earth's surface and the bottom of the ionosphere (tangential waves only). The lowest frequency, often dominant in such systems, is the fundamental mode. This fundamental frequency is given for the geometries of a spherical shell of radius $R$ or a one dimensional loop of length $L = 2\pi R$, perhaps a closed loop of transmission line (Nunez, 1995), by

$$f = \frac{gv}{L} \tag{2}$$

Here the geometric constant g is either $\sqrt{2}$ (spherical shell) or 1 (one dimensional loop). Each cortical hemisphere is topographically essentially a spherical shell. On the other hand, the postulated medium characteristic speed $v$ is the axon propagation speed in the longer systems of corticocortical axons forming in the white matter layer. Since these fibers may be substantially anisotropic with a preferred anterior-posterior orientation, it is unclear whether the shell or loop model is the most appropriate.

The wrinkled surface of each cortical hemisphere can be reshaped or mentally inflated (as with a balloon) to create an equivalent spherical shell with effective radius $R$ related to its surface area by the relation

$$R = \sqrt{\frac{A}{4\pi}} \tag{3}$$

Thus, cerebral cortex and its white matter system of (mostly) corticocortical fibers is a system somewhat analogous the earth-ionosphere shell. With a brain hemispheric surface area $A$~ 800—1500 cm$^2$ or alternately an anterior-posterior closed cortical loop of $L$~ 50—70 cm (ellipsoid-like circumference), and a characteristic corticocortical axon propagation speed of $v$~ 600—900 cm/sec (data reviewed from four independent studies in (Nunez, 1995) ), the predicted fundamental cortical frequency predicted by the naive application of Eq (2) is then

$$f \sim 8 - 26 \; \text{Hz} \tag{4}$$

We call this estimate "naive" because the fundamental mode frequency depends on both the physical shape and material properties of the wave medium (cortex-white matter). These latter properties determine the dispersive nature of the waves; that is, the precise manner in which waves distort when propagating. Such dispersive properties in cortex are expected to depend on the nature and interactions of the synaptic and action potential fields. Furthermore, cortical frequency must depend on at least one additional parameter determined by brain state. Thus, estimates in Eqs. (2) and (4) cannot be expected to represent genuine brain waves, even if the cortex were actually a spherical shell or closed loop; the postulated brain waves are much more likely to be dispersive (if for no other reason than most of Nature's waves are dispersive). Furthermore, the expected neural networks of cognitive processing (believed to be embedded in global synaptic wave fields) would be expected to cloud experimental observations of standing wave phenomenon. One may guess that such networks involve thalamocortical interactions that can generate preferred frequencies in several bands, including alpha and gamma. Thus, our scalp potentials may be viewed as some mixture of interacting global and local activity, both of which underlie and are correlated with various cognitive events.

These general ideas do not, by any stretch of the imagination, constitute a brain theory; rather they simply suggest a hypothesis and related experiments to test for *traveling* and *standing brain waves.* If estimate Eq. (4) had been obtained before the discovery of the human alpha rhythm in the 1920s, it would have provided a plausible, testable prediction. The appropriate experimental question would have been, "Can brain states be found in which neural network activity is sufficiently suppressed to allow observation of simple standing waves?" Such imagined experiments would have found the predicted EEG oscillations in the 8-13 Hz band in relaxed subjects (minimal mental load implying minimal network activity) with closed eyes (minimal visual processing).

If anything, the estimate Eq. (4) is almost *too good,* perhaps raising suspicion by critics that parameter estimates have been fudged to make a good story. But, only two parameters $v$ and $L$ are involved in the crude frequency estimate. Even if the cortical area estimate were off by a factor of two, the frequency estimate Eq. (4) would only change by $\sqrt{2}$. The axon speed estimate is based on the four independent studies reviewed in (Nunez, 1995). When PN first proposed the idea in 1972, corticocortical propagation speeds were poorly known. Axon speeds in (myelinated) peripheral axons and intracortical (non myelinated) axons are roughly ten times faster and ten times slower, respectively, than corticocortical axon speeds. That is, human axon speeds vary over at least three orders of magnitude depending mainly on axon diameter and myelination. Thus, the observed alpha frequency provided a blind prediction of corticocortical axon speed.

The simple standing brain wave model employs *Galilean idealizations* in which many essential properties of genuine brains are deliberately neglected in order to create a simple, useful model. Galileo modeled falling bodies with no air resistance even though he lacked the technology to make the air go away. Similarly, we may lack the technology to fully suppress the brain networks that might eliminate or obscure standing and traveling brain waves, although some anesthesia states may come close to this goal.

The proposed global model is based mostly on the following idea. Scalp potentials (EEG) are generated by synaptic current sources at small scales; each cubic millimeter of cortical tissue contains more than 100 million synapses. In contrast to this small scale activity, EEG data are recorded at macroscopic (centimeter) scales, thereby presenting major problems for network models attempting connections to genuine large scale data. The brain wave model follows the macroscopic dependent variables *action potential and synaptic potential densities,* for example, the number of excitatory synaptic events per square millimeter of cortical surface. All dependent variables are expressed as functions of time and cortical location. The basic approach ignores embedded network activity, although networks have been included (approximately) in more advanced models (Nunez, 1989; Jirsa & Haken, 1996). The predicted resonance frequencies for standing waves in cortex are:

$$f_n \approx \frac{v}{L}\sqrt{n^2 - (\frac{\beta \lambda L}{2\pi})^2} \qquad n = 1, 2, 3, \ldots \tag{5}$$

The symbols and estimated values are:

$v$: corticocortical propagation speed (600 - 900 cm/sec).

$L$: effective front-to-back circumference of one cortical hemisphere after inflation to a smooth surface, roughly the shape of a prolate spheroidal shell or rugby ball (50 - 70 cm).

$\lambda$: parameter indicating the fall-off in fiber density with cortical distance for the longest corticocortical fiber system (0.1 - 0.3 cm$^{-1}$).

$\beta$: nondimensional parameter controlled by neuromodulators; $\beta$ increases correspond to increased background excitability of cortex (perhaps from thalamocortical interactions, either chemical or electrical). Wave frequency and damping decrease as $\beta$ increases.

$f_n$ temporal frequencies (Hz) of fundamental mode ($n = 1$) and overtones ($n > 1$) of standing waves.

Does the theoretical *dispersion relation* Eq. (5) have any connection to genuine EEG? Surely nothing so simple can do justice to any complex brain! At best it may enjoy some approximate connections to brains in their more globally dominated states, possibly coma, anesthesia, deep sleep, some generalized epileptic states, and the more globally dominant parts of alpha rhythms. A few experimental predictions rely on this equation, but others follow only from the more general idea of standing and traveling brain waves (Nunez, 1995; Nunez, 2000; Nunez, 2010; Nunez & Srinivasan, 2006b; Nunez & Srinivasan, 2006a; Burkitt, Silberstein, Cadusch & Wood, 2000; Nunez, Wingeier & Silberstein, 2001; Posthuma, Neale, Boomsma & Geus, 2001; Wingeier, Nunez & Silberstein, 2001). Note that this model can provide only relationships not comprehensive explanations of complex physiological processes!

## 4. The Stretched String With Attached Springs

In order to distinguish theories of large-scale neocortical dynamics, we have proposed the label *local theory* to indicate mathematical models of cortical or thalamo-cortical interactions for which cortico-cortical axon propagation delays are assumed to be zero. The underlying time scales in these theories typically originate from membrane time constants giving rise to PSP rise and decay times. Thalamo-cortical networks are also "local" from the viewpoint of a surface electrode, which cannot distinguish purely cortical from thalamocortical networks. Finally, these theories are "local" in the sense of being independent of global boundary conditions dictated by the size and shape of the cortical-white matter system. By contrast, we adopt the label *global theory* to indicate mathematical models in which delays in the cortico-cortical fibers forming most of the white matter in humans provide the important underlying time scale for the large scale EEG dynamics recorded by scalp electrodes. Periodic boundary conditions are generally essential to global theories because the cortical-white matter system is topologically close to a spherical shell.

While this picture of distinct local and global models grossly oversimplifies expected genuine dynamic behaviors with substantial cross- scale interactions, it provides a convenient entry point to brain complexity. To facilitate our discussion, Figure 1 shows a stretched string with local stiffness (the little boxes) as a convenient dynamic metaphor (Nunez, 1995; Ingber, 1995a) The boxes might be simple linear springs with natural frequency $\omega_0$ or they might represent nonlinear systems organized in a complex nested hierarchy. The proposed metaphorical relationships to neocortex are outlined in Table I. String displacement is governed by the basic string equation

$$\frac{\partial^2 \Phi}{\partial t^2} - v^2 \frac{\partial^2 \Phi}{\partial x^2} + [\omega_0^2 + f(\Phi)]\Phi = 0 \qquad (6)$$

For the simple case of homogeneous linear springs attached to a homogeneous linear string of length $a$ and wave speed $v$, the normal modes of oscillation $\omega_n$ are given by

$$\omega_n^2 = \omega_0^2 + (\frac{n\pi v}{a})^2 \qquad n = 1, 2, 3, \ldots \qquad (7)$$

In this simple limiting case, the natural oscillation frequencies are seen as having distinct local and global contributions given by the first and second terms on the right side of the last equation, respectively. This same dispersion relation occurs for waves in hot plasmas and transmission lines, which might form closed loops more similar to the periodic boundary condition appropriate for neocortical standing waves. If the springs are disconnected, only the global dynamics remains. Or, if the string tension is relaxed, only the local dynamics remains. Next we approach the behavior of the nonlinear system described by the basic string equation, in which local and global effects are integrated.

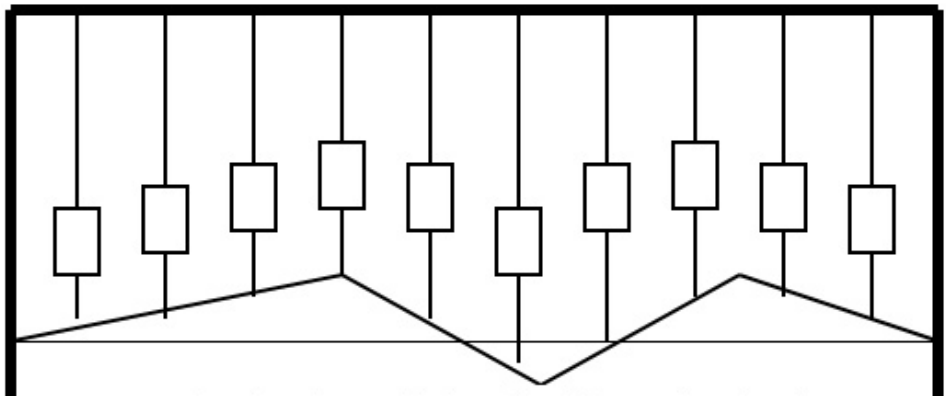

Fig. 1. The string-springs analog system. The small boxes might be simple linear springs or complex structures in a nested hierarchy analogous to columnar scale brain morphology.

| String/Spring | Neocortex/White Matter |
|---|---|
| String displacement $\Phi(x, t)$ | Any cortical field (synaptic, firing density) |
| String wave speed $v$ | Cortico-cortico axon speed |
| Spring natural frequency $\omega_0$ | Simple cortico-thalamic feedback |
| Nonlinear stiffness $\omega_0^2 + f[\Phi(x, t)]$ | Multiple-scale nonlinear columnar effects |
| Relax string tension $v \to 0$ | Ignore axon delays |
| Disconnect boxes (springs) $\omega_0, f(\Phi) \to 0$ | Ignore local dynamics |

Table I. The string-springs system as a neocortical dynamic analog

## 5. Columnar Scales

Nature has developed structures at intermediate scales in many biological as well as in many non-biological systems to facilitate flows of information between relatively small and large scales of activity. Many systems possess such structures at so-called mesoscopic scales, intermediate between microscopic and macroscopic scales, where these scales are typically defined specific to each system, and where the mesoscopic scale typically facilitates information between the microscopic and macroscopic scales. Typically, these mesoscopic scales have their own interesting dynamics.

A statistical mechanics of neocortical interactions (SMNI) for human neocortex has been developed, building from synaptic interactions to minicolumnar, macrocolumnar, and regional interactions in neocortex (Ingber, 1982; Ingber, 1983). Over a span of about 30 years, a series of about 30 SMNI papers has been developed to model columns and regions of neocortex, spanning mm to cm of tissue. SMNI uses tools of nonlinear nonequilibrium multivariate statistical mechanics, a subfield of statistical mechanics dealing with Gaussian Markovian systems with time-dependent drifts and correlated diffusions, with both drifts and diffusions nonlinear in their multiple variables.

SMNI has described columnar activity to be an effective mesoscopic scale intermediate between macroscopic regional interactions and microscopic averaged synaptic and neuronal interactions. Such treatment of neuronal activity, beyond pools of individual neurons, is based on evidence over the past 30 years of mesoscopic neocortical columnar anatomy as well as physiology which possess their own dynamics (Mountcastle, 1978; Buxhoeveden & Casanova, 2002). It is important to note that although columnar structure is ubiquitous in neocortex, it is by no means uniform nor is it so simple to define across many areas of the brain (Rakic, 2008). While SMNI has calculated phenomena like short-term memory (STM) and EEG to validate this model, there is as yet no specific real columnar data to validate SMNI's precise functional form at this scale.

When dealing with stochastic systems, there are several useful tools available when these systems can be described by Gaussian-Markovian probability distributions, even when they are in non-equilibrium, multivariate, and quite nonlinear in their means and variances. SMNI has demonstrated how most likely states described by such distributions can be calculated from the variational principle associated with systems, i.e., as Euler-Lagrange (EL) equations directly from the SMNI Lagrangian (Langouche, Roekaerts & Tirapegui, 1982). This Lagrangian is the argument in the exponent of the SMNI probability distribution. The EL equations are developed from a variational principle applied to this distribution, and they give rise to a nonlinear string model used by many neuroscientists to describe global oscillatory activity (Ingber, 1995a).

It is obvious that the mammalian brain is complex and processes information at many scales, and it has many interactions with sub-cortical structures. SMNI is appropriate to just a few scales and deals primarily with cortical structures. While SMNI has included some specific regional circuitry to address EEG calculations discussed below, details of laminar structure within minicolumns have not been included. Such laminar circuitry is of course important to many processes and, as stated in previous SMNI papers, it can be included by adding more variables. Some laminar structure is implicitly assumed in phenomena dealing with electromagnetic phenomena that depend on some systematic alignment of pyramidal neurons. Care has been taken to test SMNI at the appropriate scales, by calculating experimentally observed phenomena, and to some readers it may be surprising that it is so reasonably successful in these limited endeavors. The mathematics used is from a specialized area of multivariate nonlinear nonlinear nonequilibrium statistical mechanics (Langouche, Roekaerts & Tirapegui, 1982), and SMNI was the first physical application of these methods to the brain. In this paper, the mathematics used in all SMNI publications is not repeated, albeit referenced, but only enough mathematics is used to deal with the topic being presented.

After a short introduction to SMNI, the EL equations are presented at both regional and columnar scales. These EL equations are direct calculations of the nonlinear multivariate EL equations of the SMNI Lagrangian, giving most likely states of the system. The EL equations are quite general and are well known in physics for representing strings as well as springs, in simple as well as in complex stochastic systems, at both classical and quantum scales. This is the focus of this paper, to show how EEG may be conceptually viewed as a "string of springs."

## 6. SMNI

Neocortex has evolved to use minicolumns of neurons interacting via short-ranged interactions in macrocolumns, and interacting via long-ranged interactions across regions of macrocolumns. This common architecture processes patterns of information within and among different regions, e.g., sensory, motor, associative cortex, etc.

As depicted in Figure 2, SMNI develops three biophysical scales of neocortical interactions: (a)-$(a^*)$-(a') microscopic neurons (Sommerhoff, 1974); (b)-(b') mesocolumnar domains (Mountcastle, 1978); (c)-(c') macroscopic regions. SMNI has developed conditional probability distributions at each level, aggregating up several levels of interactions. In $(a^*)$ synaptic inter-neuronal interactions, averaged over by mesocolumns, are phenomenologically described by the mean and variance of a distribution $\Psi$ (both Poisson and Gaussian distributions were considered, giving similar results). Similarly, in (a) intraneuronal transmissions are phenomenologically described by the mean and variance of $\Gamma$ (a Gaussian distribution). Mesocolumnar averaged excitatory ($E$) and inhibitory ($I$) neuronal firings $M$ are represented in (a'). In (b) the vertical organization of minicolumns is sketched together with their horizontal stratification, yielding a physiological entity, the mesocolumn. In (b') the overlap of interacting mesocolumns at locations $r$ and $r'$ from times $t$ and $t + \tau$ is sketched. Here $\tau \sim 10$ msec represents typical periods of columnar firings. This reflects on typical individual neuronal refractory periods of ~1 msec, during which another action potential cannot be initiated, and a relative refractory period of ~ $0.5 — 10$ msec. Future research should determine which of these neuronal time scales are most dominant at the columnar time scale taken to be $\tau$. In (c) macroscopic regions of neocortex are depicted as arising from many mesocolumnar domains. (c') sketches how regions may be coupled by long–ranged interactions.

Most of these papers have dealt explicitly with calculating properties of STM and scalp EEG in order to test the basic formulation of this approach (Ingber, 1981; Ingber, 1982; Ingber, 1983; Ingber, 1984; Ingber, 1985a; Ingber, 1985b; Ingber, 1986; Ingber & Nunez, 1990; Ingber, 1991; Ingber, 1992; Ingber, 1994; Ingber & Nunez, 1995; Ingber, 1995a; Ingber, 1995b; Ingber, 1996b; Ingber, 1996a; Ingber, 1997; Ingber, 1998). The SMNI modeling of local mesocolumnar interactions, i.e., calculated to include convergence and divergence between minicolumnar and macrocolumnar interactions, was tested on STM phenomena. The SMNI modeling of macrocolumnar interactions across regions was tested on EEG phenomena.

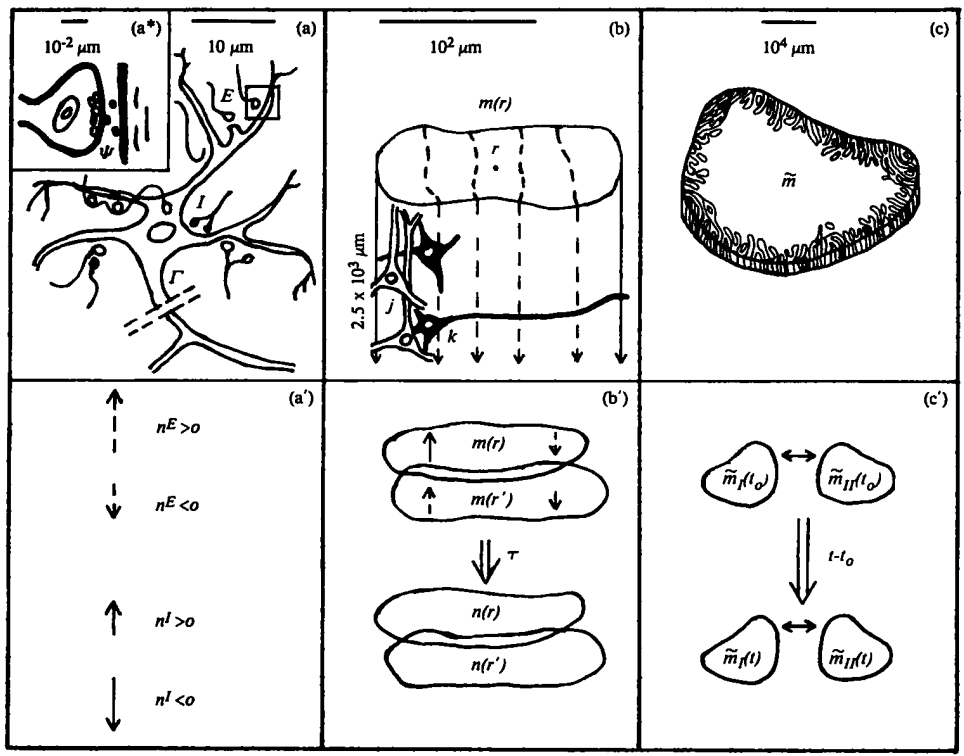

Fig. 2. Illustrated are three biophysical scales of neocortical interactions: (a)-(a*)-(a') microscopic neurons; (b)-(b') mesocolumnar domains; (c)-(c') macroscopic regions. Reprinted with permission from (Ingber, 1983) by the American Physical Society.

The EEG studies in previous SMNI applications were focused on regional scales of interactions. The STM applications were focused on columnar scales of interactions. However, this EEG study is focused at columnar scales, and it is relevant to stress the successes of this SMNI at this columnar scale, giving additional support to this SMNI model in this context. A previous report considered oscillations in quasi-linearized EL equations (Ingber, 2009a), while more recent studies consider the full nonlinear system (Ingber, 2009b).

## 6.1. SMNI STM

SMNI studies have detailed that maximal numbers of attractors lie within the physical firing space of $M^G$, where $G$ = {Excitatory, Inhibitory} = $[E, I]$ minicolumnar firings, consistent with experimentally observed capacities of auditory and visual STM, when a Centering mechanism (CM) is enforced by shifting background noise in synaptic interactions, consistent with experimental observations under conditions of selective attention (Mountcastle, Andersen & Motter, 1981; Ingber, 1984; Ingber, 1985b; Ingber, 1994; Ingber & Nunez, 1995). This leads to all attractors of the short-time distribution lying approximately along a diagonal line in $M^G$ space, effectively defining a narrow parabolic trough containing these most likely firing states. This essentially collapses the two-dimensional $M^G$ space down to a one-dimensional space of most importance. Thus, the predominant physics of STM and of (short-fiber contribution to) EEG phenomena takes place in this narrow parabolic trough in $M^G$ space, roughly along a diagonal line (Ingber, 1984).

These calculations were further supported by high-resolution evolution of the short-time conditional-probability propagator using a numerical path-integral code, PATHINT (Ingber & Nunez, 1995). SMNI correctly calculated the stability and duration of STM, propagation velocities of information across neighboring columns (Ingber, 1985a), the observed $7 \pm 2$ capacity rule of auditory memory and the observed $4 \pm 2$ capacity rule of visual memory (Ericsson & Chase, 1982; Zhang & Simon, 1985; Ingber, 1984; Ingber, 1985b), the primacy versus recency rule (Ingber, 1995b), random access to memories within tenths of a second as observed, and Hick's law of linearity of reaction time with STM information (Hick, 1952; Jensen, 1987; Ingber, 1999).

SMNI also calculates how STM patterns (e.g., from a given region or even aggregated from multiple regions) may be encoded by dynamic modification of synaptic parameters (within experimentally observed ranges) into long-term memory patterns (LTM) (Ingber, 1983).

## 6.2. Aside on STM and LTM

Most people are familiar with long-term memory storage, as recorded on their computer drives.

To illustrate dynamic short-term memory, consider an ordinary deck of cards used to produce multiple hands in some card game. Three or four normal human shuffles will typically fail to randomize the deck. As a result, each new round of hands is dealt from a deck with statistical memories created in earlier hands. In the poker game five card draw, the chances of higher cards being dealt close together will be greater than in a random deck, and this pattern will be more pronounced for the aces than for lower cards. This memory is created because players in earlier hands with (say) a pair of aces in the first betting round will have discarded three cards, keeping their aces together, even after receiving three new cards. A deck of cards can hold several distinct kinds of memory simultaneously.

## 6.3. SMNI EEG

Using the power of this formal structure, sets of EEG and evoked potential data from a separate NIH study, collected to investigate genetic predispositions to alcoholism, were fitted to an SMNI model on a lattice of regional electrodes to extract brain signatures of STM (Ingber, 1997; Ingber, 1998). Each electrode site was represented by an SMNI distribution of independent stochastic macrocolumnar-scaled $M^G$ variables, interconnected by long-ranged circuitry with delays appropriate to long-fiber communication in neocortex. The global optimization algorithm Adaptive Simulated Annealing (ASA) (Ingber, 1989; Ingber, 1993) was used to perform maximum likelihood fits of Lagrangians defined by path integrals of multivariate conditional probabilities. Canonical momenta indicators (CMI), the momentum components of the EL equations, were thereby derived for individual's EEG data. The CMI give better signal recognition than the raw data, and were used to advantage as correlates of behavioral states. In-sample data was used for training (Ingber, 1997), and out-of-sample data was used for testing (Ingber, 1998) these fits.

These results gave strong quantitative support for an accurate intuitive picture, portraying neocortical interactions as having common algebraic physics mechanisms that scale across quite disparate spatial scales and functional or behavioral phenomena, i.e., describing interactions among neurons, columns of neurons, and regional masses of neurons.

The SMNI model has been invoked to develop algorithms for better imaging resolution based on use of synchronized imaging tools, e.g., combining synchronized data from EEG, PET, SPECT, MEG, fMRI, etc (Ingber, 2009c). Other algorithms use the SMNI model to develop algorithms for artificial intelligence (Ingber, 2007; Ingber, 2008).

Below, it is reported how SMNI has calculated oscillatory behavior consistent with typical EEG, within columns, between neighboring columns via short-ranged non-myelinated fibers, across regions via myelinated fibers, and also derives a string equation consistent with the model of global EEG presented above.

## 6.4. Chaos

There are many papers on the possibility of chaos in neocortical interactions, including some that consider noise-induced interactions (Zhou & Kurths, 2003). This phenomena is useful when dealing with small networks of neurons, e.g., in some circumstances such as epilepsy. SMNI at the columnar scale can be useful to describe some forms of epilepsy, e.g., when columnar firings reach upper limits of maximal firings (Ingber, 1988).

LI took a model of chaos that might be measured by EEG, developed and published by PN and a colleague (Nunez & Srinivasan, 1993; Srinivasan & Nunez, 1993), but adding background stochastic influences and parameters that were agreed to better model neocortical interactions. The resulting multivariate nonlinear conditional probability distribution was propagated many thousands of epochs, using LI's PATHINT code, to see if chaos could exist and persist under such a model (Ingber, Srinivasan & Nunez, 1996). There was absolutely no measurable instance of chaos surviving in this more realistic context. Note that this study was at the columnar scale, not the finer scales of activity of smaller pools of neurons.

## 6.5. SMNI Application

Some of the algebra behind SMNI depicts variables and distributions that populate each representative macrocolumn in each region. While Riemannian terms were calculated when using the Stratonovich midpoint discretization of the probability distribution (Ingber, 1982; Ingber, 1983), in order to explicitly deal with the multivariate nonlinearities, here it suffices to use the more readable Ito prepoint discretization, which is an equivalent numerical distribution when used consistently (Langouche, Roekaerts & Tirapegui, 1982).

A derived mesoscopic Lagrangian $L$ defines the short-time probability distribution $P$ of firings in a minicolumn composed of ~$10^2$ neurons, where $P$ is the product of $P^G$, where $G = \{E, I\}$ chemically independent excitatory and inhibitory firing distributions, by aggregating probability distributions of neuronal firings $p_{\sigma_j}$, given its just previous interactions with all other neurons in its macrocolumnar surround. $\bar{G}$ designates contributions from both $E$ and $I$. The Einstein summation convention is used for $G$ indices, whereby repeated indices in a term implies summation over that index, unless summation is prevented by vertical bars, e.g., $|G|$. Note the use of the $\delta$-function constraint to aggregate stochastic variables, here from neuronal scales to columnar scales.

$$P = \prod_G P^G[M^G(r; t + \tau)|M^{\bar{G}}(r'; t)]$$

$$= \sum_{\sigma_j} \delta\left(\sum_{jE} \sigma_j - M^E(r; t + \tau)\right)\delta\left(\sum_{jI} \sigma_j - M^I(r; t + \tau)\right)\prod_j^N p_{\sigma_j}$$

$$\approx \prod_G (2\pi\tau g^{GG})^{-1/2} \exp(-N\tau L^G) \ ,$$

$$P \approx (2\pi\tau)^{-1/2} g^{1/2} \exp(-N\tau L) \ ,$$

$$L = L^E + L^I = (2N)^{-1}(\dot{M}^G - g^G)g_{GG'}(\dot{M}^{G'} - g^{G'}) + -V' \ ,$$

$$\dot{M}^G = [M^G(t + \tau) - M^G(t)]/\tau \ ,$$

$$V' = \sum_G V''^G_{G'}(\rho\nabla M^{G'})^2 \ ,$$

$$g^G = -\tau^{-1}(M^G + N^G \tanh F^G) \ ,$$

$$g^{GG'} = (g_{GG'})^{-1} = \delta^{G'}_G \tau^{-1} N^G \text{sech}^2 F^G \ ,$$

$$g = \det(g_{GG'}) \ ,$$

$$F^G =$$

$$\frac{(V^G - a^{|G|}_{G'} v^{|G|}_{G'} N^{G'} - \frac{1}{2} A^{|G|}_{G'} v^{|G|}_{G'} M^{G'} - a^{\ddagger E}_{E'} v^E_{E'} N^{\ddagger E'} - \frac{1}{2} A^{\ddagger E}_{E'} v^E_{E'} M^{\ddagger E'})}{((\pi/2)[(v^{|G|}_{G'})^2 + (\phi^{|G|}_{G'})^2](a^{|G|}_{G'} N^{G'} + \frac{1}{2} A^{|G|}_{G'} M^{G'} + a^{\ddagger E}_{E'} N^{\ddagger E'} + \frac{1}{2} A^{\ddagger E}_{E'} M^{\ddagger E'}))^{1/2}},$$

$$a^G_{G'} = \frac{1}{2} A^G_{G'} + B^G_{G'},$$

$$a^{\ddagger E}_{E'} = \frac{1}{2} A^{\ddagger E}_{E'} + B^{\ddagger E}_{E'}, \tag{8}$$

where $A^G_{G'}$ and $B^G_{G'}$ are minicolumnar-averaged inter-neuronal synaptic efficacies (4 combinations of $\{E, I\}$ with $\{E', I'\}$ firings), $v^G_{G'}$ and $\phi^G_{G'}$ are averaged means and variances of contributions to neuronal electric polarizations. $M^{G'}$ and $N^{G'}$ in $F^G$ are afferent macrocolumnar firings, scaled to efferent minicolumnar firings by $N/N^* \sim 10^{-3}$, where $N^*$ is the number of neurons in a macrocolumn, $\sim 10^5$. Similarly, $A^{G'}_G$ and $B^{G'}_G$ have been scaled by $N^*/N \sim 10^3$ to keep $F^G$ invariant. $V'$ are derived mesocolumnar nearest-neighbor (NN) interactions. Reasonable typical values of the postsynaptic neuronal parameters are taken to be $|v^G_{G'}| = 0.1 N^*/N$, $\phi^G_{G'} = \sqrt{0.03^2 N^*/N} = 0.173 N^*/N$. Afferent contributions from $N^{\ddagger E}$ long-ranged excitatory fibers, e.g., cortico-cortical neurons, have been added, where $N^{\ddagger E}$ might be on the order of 10% of $N^*$: Of the approximately $10^{10}$ to $10^{11}$ neocortical neurons, estimates of the number of pyramidal cells range from 1/10 to 2/3. Nearly every pyramidal cell has an axon branch that makes a cortico-cortical connection; i.e., the number of cortico-cortical fibers is of the order $10^{10}$.

Interactions via short-ranged non-myelinated fibers gives rise to the nearest-neighbor $V'$ interactions between neighboring columns (Ingber, 1982; Ingber, 1983). When EL equations associated the above Lagrangian, including $V'$ terms, are linearized, a dispersion relation was shown to support oscillatory interactions consistent with typical EEG frequencies (Ingber, 1985a). The speed of spread of information across columns, the propagation velocity in the dispersion relation, was calculated to be consistent with experimental data, e.g., observed movements of attention (Tsal, 1983) and of hallucinations (Cowan, 1982) across the visual field. Other researchers also have calculated EEG coherence across columns (Naruse, Matani, Miyawaki & Okada, 2009).

The long-ranged circuitry was parameterized (with respect to strengths and time delays) in the EEG studies described above (Ingber, 1997; Ingber, 1998). In this way SMNI presents a powerful computational tool to include both long-ranged global regional activity and short-ranged local columnar activity. This nature of physiological connectivity among columns even across regions can lead to oscillatory behavior induced among many columns.

It is interesting to note that, as originally derived (Ingber, 1982; Ingber, 1983), the numerator of $F^G$ contains information derived from presynaptic firing interactions. The location of most stable states of this SMNI system are highly dependent on the interactions presented in this numerator. The denominator of $F^G$ contains information derived from postsynaptic neuromodular and electrical processing of these firings. The nonlinearities present in this denominator dramatically affect the number and nature of stable states at scales zoomed in at magnifications on the order of a thousand times, representing neocortical processing of detailed information within a sea of stochastic activity.

## 6.6. Prototypical Cases

Three Cases of neuronal firings were considered in the first introduction of STM applications of SMNI (Ingber, 1984). Below is a short summary of these details. Note that while it suffices to define these Cases using $F^G$, the full Lagrangian and probability distribution, upon which the derivation of the EL equations are based, are themselves quite nonlinear functions of $F^G$, e.g., via hyperbolic trigonometric functions, etc.

Since STM duration is long relative to $\tau$, stationary solutions of the Lagrangian $L$, $\bar{L}$, can be investigated to determine how many stable minima $\ll \bar{M}^G \gg$ may simultaneously exist within this duration. Detailed calculations of time-dependent folding of the full time-dependent probability distribution supports persistence of these stable states within SMNI calculations of observed decay rates of STM (Ingber &

Nunez, 1995).

It is discovered that more minima of $\bar{L}$ are created, i.e., brought into the physical firing ranges, if the numerator of $F^G$ contains terms only in $\bar{M}^G$, tending to center $\bar{L}$ about $\bar{M}^G = 0$. That is, $B^G$ is modified such that the numerator of $F^G$ is transformed to

$$F'^G = \frac{-\frac{1}{2} A_{G'}^{|G|} v_{G'}^{|G|} M^{G'}}{((\pi/2)[(v_{G'}^{|G|})^2 + (\phi_{G'}^{|G|})^2](a'_{G'}^{|G|} N^{G'} + \frac{1}{2} A_{G'}^{|G|} M^{G'}))^{1/2}} \,,$$

$$a'^G_{G'} = \frac{1}{2} A_{G'}^G + B'^G_{G'} \tag{9}$$

The most likely states of the centered systems lie along diagonals in $M^G$ space, a line determined by the numerator of the threshold factor in $F^E$, essentially

$$A_E^E M^E - A_I^E M^I \approx 0 \,, \tag{10}$$

noting that in $F^I$ $I - I$ connectivity is experimentally observed to be very small relative to other pairings, so that $(A_E^I M^E - A_I^I M^I)$ is typically small only for small $M^E$.

Of course, any mechanism producing more as well as deeper minima is statistically favored. However, this particular CM has plausible support: $M^G(t+\tau) = 0$ is the state of afferent firing with highest statistical weight. I.e., there are more combinations of neuronal firings, $\sigma_j = \pm 1$, yielding this state than any other $M^G(t+\tau)$, e.g., $\sim 2^{N^G+1/2}(\pi N^G)^{-1/2}$ relative to the states $M^G = \pm N^G$. Similarly, $M^G(t)$ is the state of efferent firing with highest statistical weight. Therefore, it is natural to explore mechanisms which favor common highly weighted efferent and afferent firings in ranges consistent with favorable firing threshold factors $F^G \approx 0$.

By tuning the presynaptic stochastic background, a phenomena observed during selective attention, $F^G$ can be modified to exhibit predominately excitatory firings, inhibitory firings, or some balanced case in-between, giving rise to Cases EC, IC and BC. during selective attention, giving rise to Cases EC, IC and BC. It is observed that visual neocortex has twice the number of neurons per minicolumn as other regions of neocortex. In the SMNI model this gives rise to fewer and deeper STM states, consistent with the observed $4 \pm 2$ capacity rule of these memory states. These calculations are Cases ECV, ICV and BCV. The details of such calculations are given in previous SMNI papers (Ingber, 1984; Ingber, 1985b; Ingber, 2009b).

## 7. Euler-Lagrange (EL)

The EL equations are derived from the long-time conditional probability distribution of columnar firings over all cortex, represented by $\bar{M}$, in terms of the Action $S$, The path integral has a variational principle, $\delta L = 0$ which gives the EL equations for SMNI (Ingber, 1982; Ingber, 1983).

When dealing when multivariate Gaussian stochastic systems with nonlinear drifts and diffusions, it is possible to work with three essentially mathematically equivalent representations of the same physics: Langevin equations — coupled stochastic differential equations, a Fokker-Plank equation — a multivariate partial differential equation, and a path-integral Lagrangian — detailing the evolution of the short-time conditional probability distribution of the variables (Langouche, Roekaerts & Tirapegui, 1982).

While it typically takes more numerical and algebraic expertise to deal with the path-integral Lagrangian, there are many benefits, including intuitive numerical and algebraic tools. For example, the Lagrangian components and EL equations are essentially the counterpart to classical dynamics,

$$\text{Mass} = g_{GG'} = \frac{\partial^2 L}{\partial(\partial M^G/\partial t)\partial(\partial M^{G'}/\partial t)} \,,$$

$$\text{Momentum} = \Pi^G = \frac{\partial L}{\partial(\partial M^G/\partial t)} \,,$$

$$\text{Force} = \frac{\partial L}{\partial M^G} \, ,$$

$$\text{F} - \text{ma} = 0: \ \ \delta L = 0 = \frac{\partial L}{\partial M^G} - \frac{\partial}{\partial t} \frac{\partial L}{\partial(\partial M^G / \partial t)} \tag{11}$$

The most-probable firing states derived variationally from the path-integral Lagrangian as the EL equations represent a reasonable average over the noise in the SMNI system. For many studies, the noise cannot be simply disregarded, as demonstrated in other SMNI STM and EEG studies, but for the purpose here of demonstrating the existence of multiple local oscillatory states that can be identified with EEG frequencies, the EL equations serve very well.

The Lagrangian and associated EL equations have been developed at SMNI columnar scales, as well as for regional scalp EEG activity by scaling up from the SMNI columnar scales as outlined below.

## 7.1. Strings

The nonlinear string model was derived using the EL equation for the electric potential $\Phi$ measured by EEG, considering one firing variable along the parabolic trough of attractor states being proportional to $\Phi$ (Ingber & Nunez, 1990).

Since only one variable, the electric potential is being measured, is reasonable to assume that a single independent firing variable might offer a crude description of this physics. Furthermore, the scalp potential $\Phi$ can be considered to be a function of this firing variable. (Here, "potential" refers to the electric potential, not any potential term in the SMNI Lagrangian.) In an abbreviated notation subscripting the time-dependence,

$$\Phi_t - \ll \Phi \gg = \Phi(M_t^E, M_t^I) \approx a(M_t^E - \ll M^E \gg) + b(M_t^I - \ll M^I \gg) \, , \tag{12}$$

where $a$ and $b$ are constants, and $\ll \Phi \gg$ and $\ll M^G \gg$ represent typical minima in the trough. In the context of fitting data to the dynamic variables, there are three effective constants, $\{ a, b, \phi \}$,

$$\Phi_t - \phi = a M_t^E + b M_t^I \tag{13}$$

We scale and aggregate the mesoscopic columnar probability distributions, $P$, over this columnar firing space to obtain the macroscopic conditional probability distribution over the scalp-potential space:

$$P_\Phi[\Phi] = \int dM^E dM^I P[M^E, M^I] \delta[\Phi - \Phi'(M^E, M^I)] \tag{14}$$

Note that again we use the $\delta$-function constraint to aggregate stochastic variables, here columnar firing states into electric potentials. The parabolic trough described above justifies a form

$$P_\Phi = (2\pi\sigma^2)^{-1/2} \exp(-\Delta t \int dx \, L_\Phi) \, ,$$

$$L_\Phi = \frac{\alpha}{2} |\partial\Phi/\partial t|^2 + \frac{\beta}{2} |\partial\Phi/\partial x|^2 + \frac{\gamma}{2} |\Phi|^2 + F(\Phi) \, ,$$

$$\sigma^2 = 2\Delta t / \alpha \, , \tag{15}$$

where $F(\Phi)$ contains nonlinearities away from the trough, $\sigma^2$ is on the order of $1/N$ given the derivation of $\underline{L}$ above, and the integral over $x$ is taken over the spatial region of interest. In general, there also will be terms linear in $\partial\Phi/\partial t$ and in $\partial\Phi/\partial x$.

Here, the EL equation includes variation across the spatial extent, $x$, of columns in regions,

$$\frac{\partial}{\partial t} \frac{\partial L}{\partial(\partial\Phi/\partial t)} + \frac{\partial}{\partial x} \frac{\partial L}{\partial(\partial\Phi/\partial x)} - \frac{\partial L}{\partial\Phi} = 0 \tag{16}$$

The result is

$$\alpha \frac{\partial^2 \Phi}{\partial t^2} + \beta \frac{\partial^2 \Phi}{\partial x^2} + \gamma \Phi - \frac{\partial F}{\partial \Phi} = 0 \tag{17}$$

The determinant prefactor $g$ defined above also contains nonlinear details affecting the state of the system. Since $g$ is often a small number, distortion of the scale of $L$ is avoided by normalizing $g/g_0$, where $g_0$ is simply $g$ evaluated at $M^E = M^{\ddagger E'} = M^I = 0$.

If there exist regions in neocortical parameter space such that we can identify $\beta/\alpha = -c^2$, $\gamma/\alpha = \omega_0^2$ (e.g., as explicitly calculated using the CM),

$$\frac{1}{\alpha} \frac{\partial F}{\partial \Phi} = -\Phi f(\Phi) , \tag{18}$$

then we recover the nonlinear string model.

Note that this string derivation is only consistent with the global string analog described in the first sections of this paper, if the spatial extent is extended across the scalp via long-ranged fibers connecting columns with $M^{\ddagger E'}$ firings. This leads to a string of columns. The next section calculates an EL model that shows that these columns can be viewed as spring analogs.

## 7.2. Springs

For a given column in terms of the probability description given above, the above EL equations are represented as

$$\frac{\partial}{\partial t} \frac{\partial L}{\partial(\partial M^E/\partial t)} - \frac{\partial L}{\partial M^E} = 0 ,$$

$$\frac{\partial}{\partial t} \frac{\partial L}{\partial(\partial M^I/\partial t)} - \frac{\partial L}{\partial M^I} = 0 \tag{19}$$

To investigate dynamics of multivariate stochastic nonlinear systems, such as neocortex presents, it is not sensible to simply apply simple mean-field theories which assume sharply peaked distributions, since the dynamics of nonlinear diffusions in particular are typically washed out.

Previous SMNI EEG studies had demonstrated that simple linearized dispersion relations derived from the EL equations support the local generation of frequencies observed experimentally as well as deriving diffusive propagation velocities of information across minicolumns consistent with other experimental studies. The earliest studies simply used a driving force $J_G M^G$ in the Lagrangian to model long-ranged interactions among fibers (Ingber, 1982; Ingber, 1983). Subsequent studies considered regional interactions driving localized columnar activity within these regions (Ingber, 1996b; Ingber, 1997; Ingber, 1998).

A recent set of calculations examined these columnar EL equations to see if EEG oscillatory behavior could be supported at just this columnar scale, i.e., within a single column. At first, the EL equations were quasi-linearized, by extracting coefficients of $M$ and $dM/dt$. The nonlinear coefficients were presented as graphs over all firing states (Ingber, 2009a). This exercise demonstrated that a spring-type model of oscillations was plausible. Then a more detailed study was performed, developing over two million lines of C code from the algebra generated by an algebraic tool, Maxima, to see what range of oscillatory behavior could be considered as optimal solutions satisfying the EL equations (Ingber, 2009b). The answer was affirmative, in that ranges of $\omega t \approx 1$ were supported, implying that oscillatory solutions might be sustainable just due to columnar dynamics at that scale. Below, the full probability distribution is evolved with such oscillatory states, confirming this is true.

To understand the nature of the EL equations, it is useful to view the probability space over which most likely states exist. Figure 3 presents the probability distribution over firing space, as the PATHINT code evolves it over 1 sec, folding every $\tau/10$, or 1000 folding to reach 1 sec. Similar to the previous calculations (Ingber & Nunez, 1995), to control large negative drifts at the boundaries which can cause anomalous numerical problems at the edges of firing space, a simple Gaussian cutoff with width 0.1 was taken for the drifts at those boundaries. This cutoff was applied as well to the diffusions, which kept the mesh uniform near the edges, resulting in the over all mesh being within 1-3 (2-4) firing units for non-visual (visual) Cases. This coarse mesh results in some fine structure in the trough not visible in the graphs, but the overall structure of the distributions are clear. Each folding took about 0.045 (0.16) secs on a dedicated IBM a31p Thinkpad with 1 GB RAM running at 1.8 GHz for non-visual (visual) Cases, under gcc/g++-4.3.3 under Linux Ubuntu 9.04. The kernel is a banded matrix with over 250K (500K) non-zero entries for non-visual (visual) Cases. It is clear that columnar STM under all 4 Cases is quite stable for at least 1 sec. Similar results are obtained for Cases BCV, EC and IC.

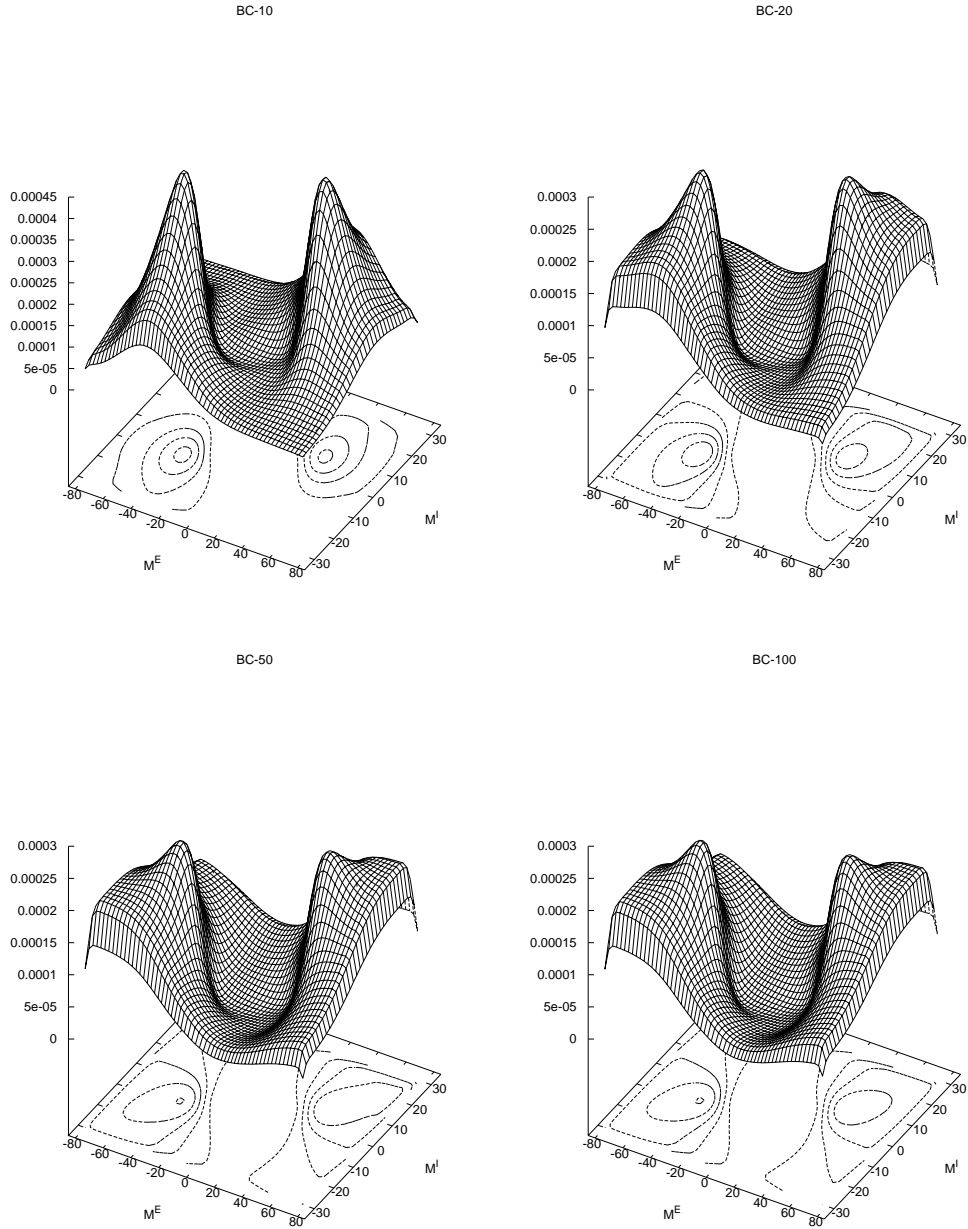

Fig. 3. The probability distribution over firing space for case BC is shown as it develops at times $10\tau$, $20\tau$, $50\tau$, and $100\tau$.

Figure 4 presents the probability distribution over firing space for case BCosc, with explicit driving oscillatory firing states by setting $M^G \to M^G \cos(\omega_G t)$ with $\omega_G = 1$. The time-dependent folding takes about 10 secs per iteration on the same Ubuntu machine due to normalization conditions on columns of the kernel. STM still persists for over 1/2 sec, but fades by 1 sec. Similar results are obtained for Cases BCV, EC and IC.

Of course, the effects of oscillatory factors will be much greater when $\omega_G t$ reaches phases such that these factors pass through zero. Here, the maximum phase reaches 1 with cos chosen as the factor.

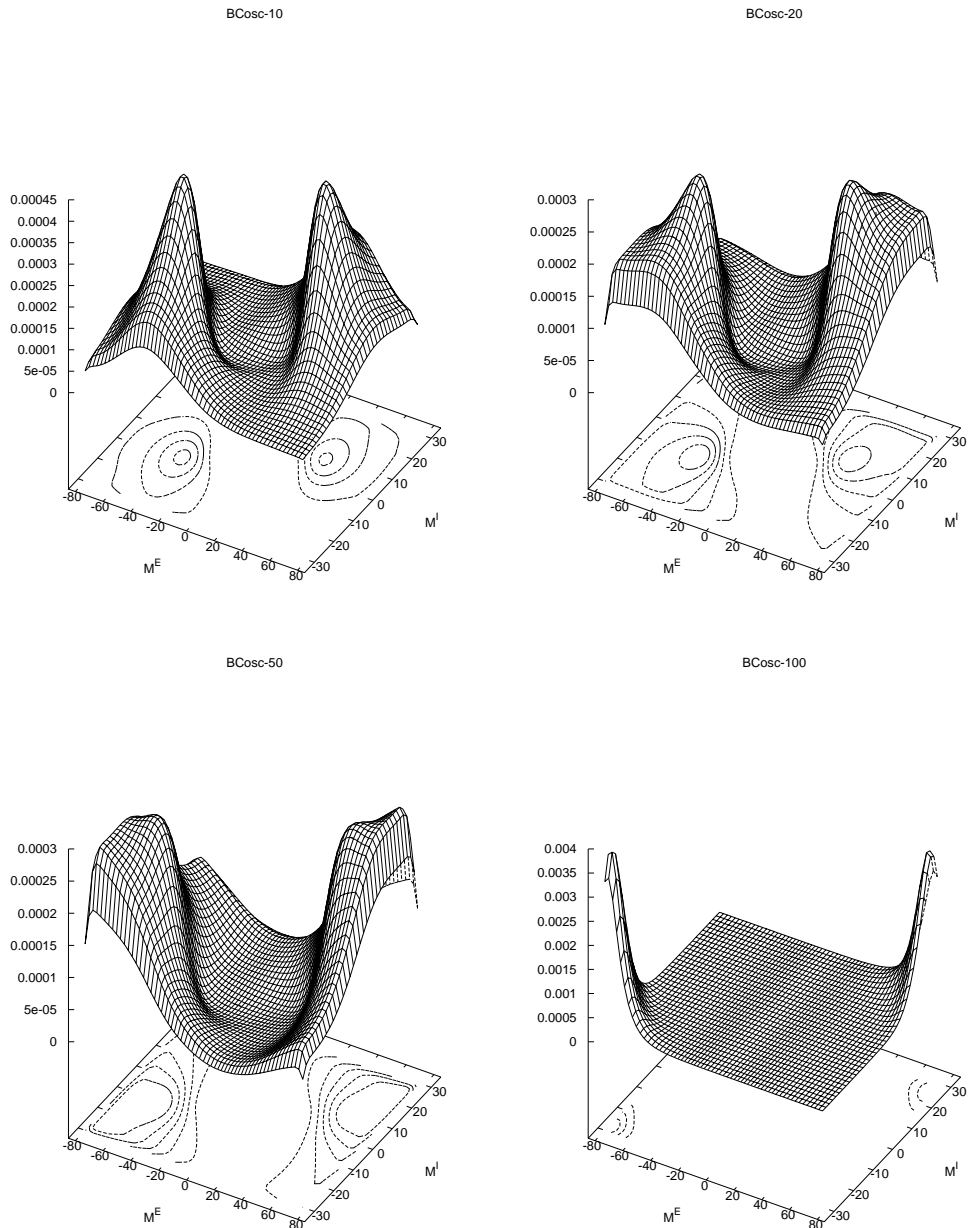

Fig. 4. The probability distribution over oscillatory firing space is shown as it develops at times $10\tau$, $20\tau$, $50\tau$, and $100\tau$, with explicit oscillatory firing states.

Figure 5 presents the probability distribution over firing space for case BCreg, with driving oscillatory firing states from an external column, e.g., from a distant regions, here set to $M^{\ddagger E'} = 10\cos(\omega_E t)$, where a weight of 10% if given to these external firings. Since the oscillatory forces are from a small percentage of all neurons represented in the column, e.g., as compared to the case in Figure 4, the changes are smaller but noticeable. Similar results are obtained for Cases BCV, EC and IC.

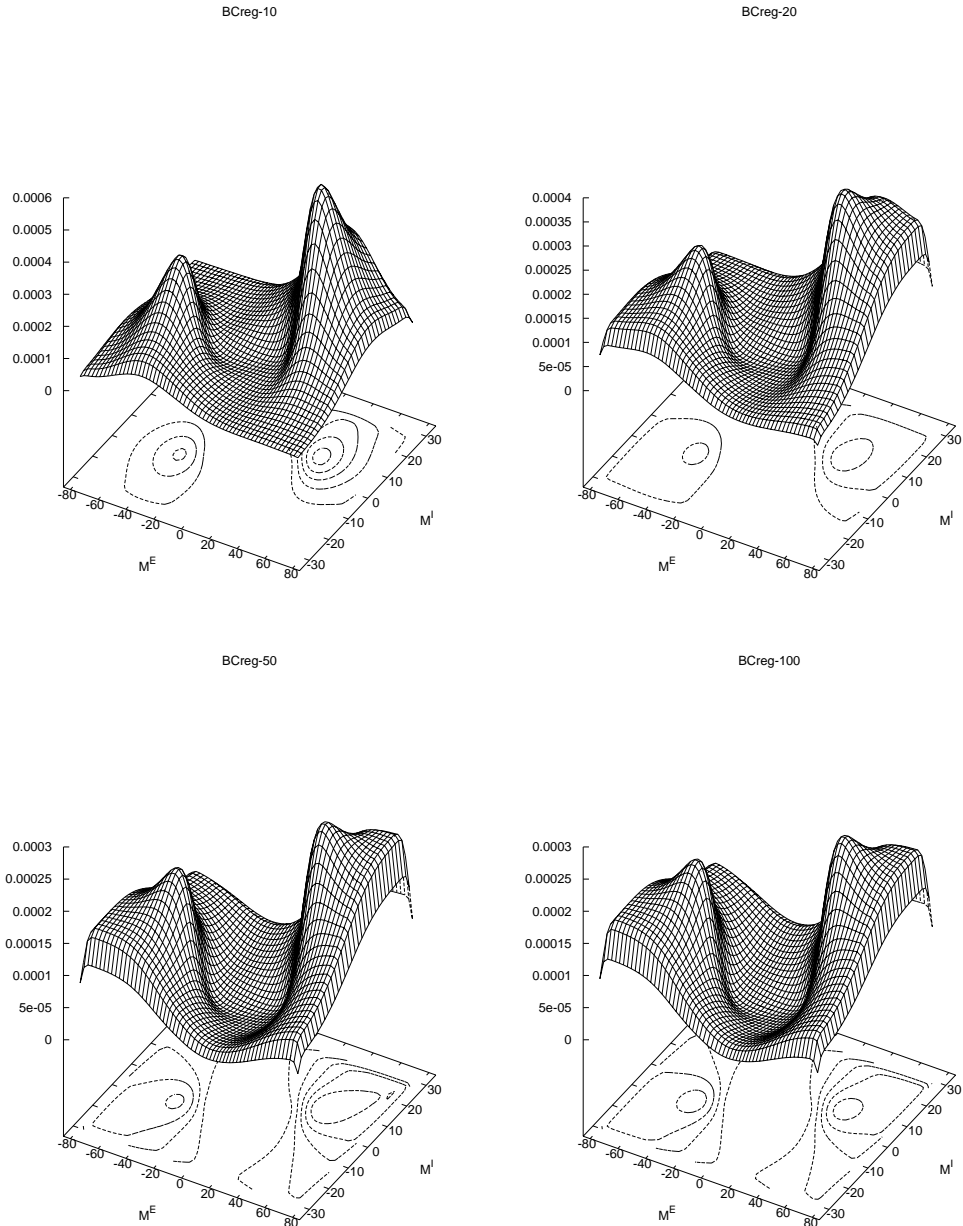

Fig. 5. The probability distribution driven by external oscillatory firings is shown as it develops at times $10\tau$, $20\tau$, $50\tau$, and $100\tau$.

## 8. Driven by noise

Using SMNI, scenarios mentioned above can be detailed. For example, if oscillatory behavior is generated within a given column — especially a column with the CM on, then these oscillations may be induced in other columns — especially those with the CM on and with which it has strong connectivity via long-ranged $M^{\ddagger E}$ firings which contribute to their local threshold factors $F^G$. Therefore it is reasonable to conjecture that if columnar firings of short-ranged fibers $M^G$ can oscillate within ranges of oscillations of long-ranged fibers $M^{\ddagger E}$, this could facilitate information processed at fine neuronal and synaptic scales to be carried across minicolumns and regional columns with relative efficiency. Note that this activity is at levels of $10^{-2}$ or $10^{-3}$ of the Lagrangian defining a small scale for STM, i.e., zooming in to still within classical (not quantum) domains of information, e.g., at the scale being sensitive to one to several neurons.

While attractor states have been explicitly detailed in previous papers for several SMNI models, here oscillatory states have been calculated throughout the range of firing space. Given that long-ranged fiber interactions across regions can constrain columnar firings, it is useful to at least learn how oscillations may be supported in limited ranges of such constrained firings.

The results show that only under conditions suitable for STM do columnar interactions per se support spectra of oscillatory behavior $\omega_G$ in observed frequency ranges robust throughout $M^G$ firing space. In retrospect, this is not too surprising, since some coherent interactions are likely required to sustain multiple stable states for STM. This leads to a strong inference that physiological states of columnar activity receptive to selective attention support oscillatory processing in these ranges. Note that selective attention even to information processed within a given region of neocortex likely requires interactions with frontal cortex and/or sub-cortical structures not explicitly included in the SMNI model.

The sensitivity of stochastic multivariate nonlinear (multiple quasi-stable states) to relatively weak oscillatory forces has been documented in many systems (Lindner, Garcia-Ojalvo, Neiman & Schimansky-Geier, 2004). Stochastic resonance has been demonstrated in mammalian brain, using relatively weak electric fields to effect sinusoidal signals in stochastic firings of groups of neurons (Gluckman, Netoff $et\ al$, 1996). In SMNI, noise arises at synaptic levels, and the sensitivity at issue in STM is at the aggregated mesoscopic level of columns of neuronal distributions. The averaged synaptic noise is a parameter which appears in the mean as well as the covariance of the aggregated system via the threshold factors $F^G$. As introduced here at the columnar level, oscillatory changes in firings within the duration of STM shifts the stable STM states in firing space, directly affecting access to these states.

The source of the background synaptic noise, especially presynaptic noise which gives rise to the CM, also is a long-standing area of research (Gluckman, Netoff $et\ al$, 1996). Further research into the roles of the CM and columnar support for EEG, together with other proposed mechanisms for columnar-glial magnetic interactions for some control of glial-presynaptic background interactions, includes a path for future investigations outlined above to test for quantum-classical interactions that directly support STM by controlling presynaptic noise.

STM (or working memory), along with selective (or focused) attention to this memory, are generally considered important aspects of the "easy" problem of consciousness, e.g., where objective neural correlates of consciousness (NCC) are sought, without addressing the "hard" aspects of subjective and phenomenal states, e.g., "qualia" (Crick & Koch, 1998). In the absence of selective attention, unconscious processing of information and computation can still take place using STM. In this context, such research in consciousness and unconscious information processing must include the dynamics of STM.

## 9. Summary and Conclusion

We have suggested that dynamic behavior in neocortex is due to some combination of global and local processes with important top-down and bottom-up interactions across spatial scales, a typical feature of many if not most complex physical, biological, social, financial, and other systems. We have focused on electroencephalography (EEG) because EEG provides most of the existing data on the relationship between ms scale neocortical dynamics and brain state. Although EEG recorded from the human scalp provides data at very large spatial scales (several cm), it is closely correlated with many distinct kinds of cognitive processing.

A purely global EEG model stresses myelinated axon propagation delays and periodic boundary conditions in the cortical-white matter system. As this system is topologically close to a spherical shell, standing waves are predicted with fundamental frequency in the typical EEG range near 10 Hz. In sharp contrast to the purely global model, the proposed local mechanisms are multiscale interactions between cortical columns via short-ranged non-myelinated fibers. A statistical mechanics of neocortical interactions (SMNI) predicts oscillatory behavior within columns, between neighboring columns and via short-ranged non-myelinated fibers. The columnar dynamics, based partly on membrane time constants, also predicts frequencies in the range of EEG.

We generally expect both local and global processes to influence EEG at all scales, including the large scale scalp data. Thus, SMNI also includes interactions across cortical regions via myelinated fibers effecting coupling the local and global models. The combined local-global dynamics is demonstrated with an analog mechanical system consisting of a stretch string (producing standing waves) with attached nonlinear springs representing columnar dynamics. SMNI is able to derive a string equation consistent with the global EEG model. We conclude that the string-spring system provides an excellent analog with several general features that parallel multiscale interactions in genuine neocortex.